\documentclass[lettersize,journal]{IEEEtran}
\usepackage{amsmath,amsfonts}
\usepackage{array}
\usepackage[caption=false,font=normalsize,labelfont=sf,textfont=sf]{subfig}
\usepackage{textcomp}
\usepackage{stfloats}
\usepackage{url}
\usepackage{verbatim}
\usepackage{graphicx}
\usepackage{cite}
\usepackage[ruled,vlined]{algorithm2e}
\usepackage{tabularx}
\usepackage{makecell}
\usepackage{balance}
\usepackage{multirow} 
\usepackage{listings}
\usepackage{xcolor}
\usepackage[T1]{fontenc}
\usepackage{orcidlink}
\usepackage{pifont} 

\begin{document}

\title{DySec: A Machine Learning-based Dynamic Analysis for Detecting Malicious Packages in PyPI Ecosystem}

\author{Sk~Tanzir~Mehedi~\orcidlink{0000-0003-4435-7856}, Chadni~Islam~\orcidlink{0000-0002-6349-6483}, Gowri~Ramachandran~\orcidlink{0000-0001-5944-1335}, and Raja~Jurdak~\orcidlink{0000-0001-7517-0782}

\thanks{(Corresponding author: Sk Tanzir Mehedi)}
\thanks{Sk Tanzir Mehedi, Gowri Ramachandran, and Raja Jurdak are with the Queensland University of Technology, Brisbane, QLD 4000, Australia (e-mail: tanzir.mehedi@hdr.qut.edu.au; {\{g.ramachandran, r.jurdak}\}@qut.edu.au); Chadni Islam is with the Edith Cowan University, Joondalup, WA 6027, Australia (e-mail: c.islam@ecu.edu.au)}
}



\maketitle

\begin{abstract}
Malicious Python packages make software supply chains vulnerable by exploiting trust in open-source repositories like Python Package Index (PyPI). Lack of real-time behavioral monitoring makes metadata inspection and static code analysis inadequate against advanced attack strategies such as typosquatting, covert remote access activation, and dynamic payload generation. To address these challenges, we introduce DySec, a machine learning (ML)-based dynamic analysis framework for PyPI that uses eBPF kernel and user-level probes to monitor behaviors during package installation. By capturing 36 real-time features—including system calls, network traffic, resource usage, directory access, and installation patterns—DySec detects threats like typosquatting, covert remote access activation, dynamic payload generation, and multiphase attack malware. We developed a comprehensive dataset of 14,271 Python packages, including 7,127 malicious sample traces, by executing them in a controlled isolated environment. Experimental results demonstrate that DySec achieves a 95.99\% detection accuracy with a latency of <0.5s, reducing false negatives by 78.65\% compared to static analysis and 82.24\% compared to metadata analysis. During the evaluation, DySec flagged 11 packages that PyPI classified as benign. A manual analysis, including installation behavior inspection, confirmed six of them as malicious. These findings were reported to PyPI maintainers, resulting in the removal of four packages. DySec bridges the gap between reactive traditional methods and proactive, scalable threat mitigation in open-source ecosystems by uniquely detecting malicious install-time behaviors.
\end{abstract}


\section{Introduction}
\IEEEPARstart{O}{pen-source} software has become an integral part of the software supply chain. According to the Open Source Security and Risk Analysis (OSSRA) report of 2024, 96\% of the 1,067 codebases scanned contain open source code, while 77\% of all code in the total codebases originates from open source code~\cite{ossra2024}. However, its openness and decentralized structure add hurdles to ensuring the privacy, security, and reliability of the software supply chain~\cite{HIPAAJournal2025, 9740718}. These software supply chain challenges are particularly evident in the ecosystem of open source packages, the modular components, that power much of today’s software~\cite{ossra2024, hybrid_samaana_2024}. In August 2024, more than 704,102 malicious components were identified, a 156\% year-over-year increase. A significant manifestation of these challenges can be observed within the Python Package Index (PyPI) ecosystem, an official third-party software repository for Python~\cite{jfrog2024, sonatype2024}. 

\begin{figure}[htp!]
    \centering
    \includegraphics[width=0.92\linewidth]{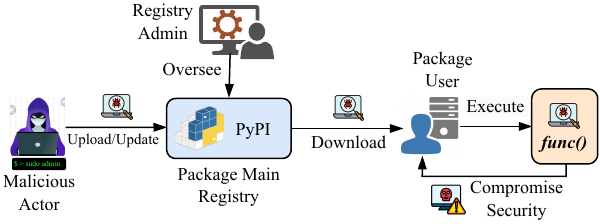}
    \caption{Threats in the PyPI package registry ecosystem.}
    \label{fig:Overall workflow}
\end{figure}

As of December 2024, PyPI hosted more than 590,500 packages and facilitated 1.882 billion daily downloads, serving as a cornerstone of Python’s open-source ecosystem~\cite{pypi, pypistats, depsdev}. Its extensive collection of libraries and tools empowers developers, data scientists, researchers, and businesses, driving innovation across various industries, from web development and data science to machine learning and scientific computing~\cite{pypistats}. The health and accessibility of PyPI are therefore crucial to the continued growth and success of Python and its broad user base~\cite{ruohonen2021large}. However, its accessibility and scale have made it a prime target for malicious actors~\cite{hybrid_samaana_2024}. Recent reports up to July 2024 have identified 7,127 PyPI packages (1.2\%) as malicious~\cite{snyksecurity, nvdnist, VirusTotal, blackduck2024}. The malicious packages caused different types of attacks such as data exfiltration using AWS keys or remote API, credential theft through typosquatting, and remote code execution via dependency confusion attacks~\cite{phishing_ori_2023, hackernews2024, checkmarx2024}. For instance, the `Zebo-0.1.0' package periodically captures screenshots and uploads them to an attacker-controlled server using a remote API~\cite{fortinet2025}. This exemplifies the broader threats within the PyPI package registry ecosystem, where users may accidentally install malicious packages. Figure~\ref{fig:Overall workflow} illustrates how attackers employ sophisticated techniques to evade security defenses. These evolving attack strategies highlight the need for robust security measures to detect and mitigate malicious packages, ensuring the integrity of the PyPI ecosystem.

\vspace{-5pt}

\begin{figure}[htp!]
    \centering
    \includegraphics[width=0.92\linewidth]{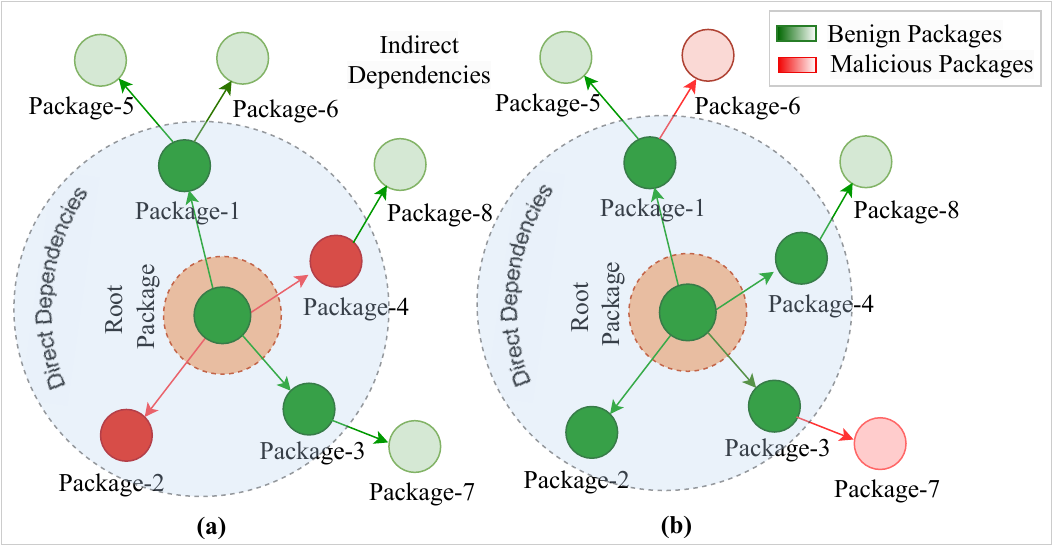}
    \caption{Dependency graphs showing (a) direct malicious dependencies and (b) indirect malicious dependencies.}
    \label{fig:Dependencies graph}
\end{figure}

Existing studies aim to enhance the detection of malicious PyPI packages through various methods, including metadata analysis~\cite{metadata_halder_2024, metadata_bommarito_2019}, static analysis~\cite{hybrid_samaana_2024, static_ruohonen_2021, static_zhang_2023}, and hybrid analysis~\cite{hybrid_damodaran_2022, hybrid_afianian_2020}. However, these methods face critical limitations in addressing evolving threats, such as detecting typosquatting, covert remote access activation, and dynamic payload generation~\cite{zheng2024robust, phpass2024, martini2019python3dateutil}. Metadata analysis, which relies on package details, is easily bypassed by attackers using fake credentials to mimic legitimate packages. For instance, the malicious package `python3-dateutil' deceived users by impersonating a trusted library, leading to the installation of a trojanized version that stole passwords and authentication tokens~\cite{martini2019python3dateutil}. Static analysis, which inspects code without execution, struggles to detect threats due to typosquatting, remote execution, dynamically generated payloads, and multiphase attack sequences during installation—issues that often result in high false positive rates~\cite{static_limt_moser_2007, false_positive_lungana_2018}. For example, `phpass', which evades detection by generating malicious payloads at install-time using encoded strings; once deployed, it exfiltrates environment variables such as AWS access keys to an attacker-controlled server~\cite{phpass2024}. Hybrid analysis, while more accurate than static or metadata methods, remains vulnerable to sandbox-aware malware that delays execution until after analysis. The package `Pymafka' exploited this limitation by activating a keylogger immediately after installation, bypassing sandbox checks, and stealing sensitive device data~\cite{sharma2022pymafka}. Furthermore, all detection methods struggle to identify multiphase threats hidden in nested dependency chains, making them ineffective against indirect malicious dependencies~\cite{dependency_Snyk_2024}. Figure~\ref{fig:Dependencies graph} highlights this challenge by contrasting dependency graphs of benign and malicious packages. While direct malicious dependencies~(a) are detectable via existing methods, indirect dependencies~(b) remain hidden within packages, exploiting install-time activation to evade detection. 

\vspace{-5pt}

\vspace{-5pt}

To address the limitations of existing detection methods inability to detect typosquatting, covert remote access activation, dynamic payload generation, and multiphase attack sequences, we propose an ML-based \textbf{Dy}namic \textbf{Sec}urity Analysis framework \textbf{(DySec)} for the PyPI ecosystem. DySec employs the extended Berkeley Packet Filter (eBPF) kernel and user-level probes to monitor behaviors during package installation~\cite{ebpf_eitani_2023, ebpf_bogle_2023, ebpf_jia_2023}. Unlike traditional tools such as Wireshark and Sysdig, eBPF enables lightweight, real-time monitoring without requiring kernel modifications~\cite{ebpf_Zhuravchak_2023}. Moreover, eBPF’s programmability in C or Python allows DySec to adapt dynamically to evolving security threats~\cite{ebpf_eitani_2023}. To effectively detect malicious PyPI packages, DySec extracts dynamic features, including system calls, network traffic, resource usage, directory access, and installation patterns. It also analyzes behavioral patterns exhibited during package installation to identify anomalies that indicate stealthy and install-time attack behaviors. To evaluate DySec’s effectiveness, we created a dataset of real-world malicious PyPI package traces. These packages were executed in a controlled, isolated environment. DySec's performance was then compared against existing detection methods. The package list we considered is publicly available on DySec.io\footnote{\url{https://dysec.io}}, ensuring reproducibility and further research. The key contributions include:

\begin{itemize}
    \item An ML-based dynamic analysis framework, DySec detects malicious PyPI packages by monitoring install-time behavior using eBPF-based kernel and user-level probes.
    
    \item A dataset of 14,271 packages, including 7,127 malicious ones, with 36 features across six categories that capture install-time behaviors previously unexplored for malicious package detection. The dataset will be publicly available upon publication to support future research.

    \item A comprehensive analysis of DySec's performance demonstrates a detection accuracy of 95.99\% with a latency of <0.5s, reducing false negatives by 78.65\% compared to static analysis and 82.24\% compared to metadata analysis, leading to the removal of four malicious packages from PyPI.
    
\end{itemize}
    
This study is structured as follows: Section~\ref{Related Work} reviews existing approaches for detecting malicious packages. Section~\ref{Methodology} details DySec’s framework, while Section~\ref{Experimental Design and Setup} covers the dataset, experimental design, and implementation. Section~\ref{Evaluation} presents the results,  and Section~\ref{Threats to Validity} discusses the threats to validity. Finally, Section~\ref{Conclusion} concludes the paper.

\vspace{-5pt}
\section{Related Work}
\label{Related Work}

This section provides an overview of recent approaches to malicious package detection, including metadata analysis, static analysis, hybrid techniques, and eBPF-based methods, before introducing the proposed methodology.
    
\textbf{Metadata-based detection:} Malicious package detection using metadata has gained attraction due to its efficiency and scalability~\cite{metadata_bommarito_2019, metadata_gao_2024, metadata_halder_2024}. Halder et al.~\cite{metadata_halder_2024} demonstrated effectiveness of metadata in identifying malicious PyPI packages, achieving high precision with ML models. Examples of metadata-based features include package descriptions, versioning patterns, and author profiles. Earlier work by Leckie and Yasinsac~\cite{metadata_leckie_2004} established the utility of metadata for anomaly detection in security protocols, emphasizing its role in attack deduction. Mutmbak et al.~\cite{metadata_mutmbak_2022} extended this to network security, using metadata like traffic patterns for anomaly detection. For software ecosystems, Bommarito and Bommarito~\cite{metadata_bommarito_2019} analyzed structural patterns in PyPI metadata, while Ohm et al.~\cite{metadata_ohm_2022} identified metadata such as package naming conventions for detecting malicious npm packages. Besides research, there is a growing rise in tools designed to assess the trustworthiness of PyPI packages. For instance, PyRadar~\cite{metadata_gao_2024}, Pyroma~\cite{metadata_pyroma}, and Twine~\cite{metadata_twine} are proposed to evaluate and validate Python package metadata, ensuring compliance with best practices and distribution standards. 

Despite its effectiveness, the ease of altering metadata undermines the reliability of metadata-based analysis. Attackers can easily manipulate metadata to introduce malicious payloads in various ways, such as falsifying maintainer details or hijacking abandoned packages~\cite{hybrid_samaana_2024}.

\textbf{Static analysis-based detection:} Recent research trends have highlighted the increasing use of static code analysis for malware detection in package ecosystems such as npm and PyPI~\cite{static_guo_2023, static_sejfia_2022}. For instance, Sejfia and Schäfer~\cite{static_sejfia_2022} developed an automated framework for npm packages using AST-based pattern matching and heuristics to detect malicious code snippets. Similarly, a set of studies has used static code analysis to identify vulnerabilities in PyPI packages. Ruohonen et al.~\cite{static_ruohonen_2021} conducted a large-scale analysis detecting malicious patterns and vulnerabilities through code inspection, while Guo et al.~\cite{static_guo_2023} empirically analyzed malicious PyPI packages, uncovering common obfuscation techniques and payload delivery mechanisms. Zhang et al.~\cite{static_zhang_2023} proposed a unified model for detecting malicious packages in npm and PyPI by analyzing behavioral sequences in code. However, Moser et al.~\cite{static_limt_moser_2007} cautioned that static analysis alone has inherent limitations, such as the evasion of detection mechanisms through dynamic code generation and multiphase attack sequences. Tools like PyPI Malware Checks (PyPA)~\cite{malwarechecks}, OSSGadget (Microsoft)~\cite{ossgadget}, and Bandit4Mal (Vu)~\cite{bandit4mal} enhance traditional tools like Bandit~\cite{bandit} by adding specialized rules for detecting malicious code. However, their effectiveness relies on predefined patterns, leading to increased false positives, and they are unable to detect install-time payload generation or multiphase attack sequences~\cite{ruohonen2021large, static_limt_moser_2007, zheng2024robust}. Despite their utility, static tools struggle with false positive rate, typosquatting, covert remote access activation, dynamic payload generation, and multiphase attack sequences ~\cite{zheng2024robust, false_positive_lungana_2018}. As a result, researchers have increasingly turned to hybrid analysis to overcome these limitations.

\textbf{Hybrid model-based detection:} Recent research trends also highlight the growing use of hybrid analysis for malware detection in npm and PyPI~\cite{hybrid_samaana_2024, hybrid_afianian_2020}. This method integrates metadata and static analysis to address limitations such as false positive rates and obfuscated code. For instance, Samaana et al.~\cite{hybrid_samaana_2024} integrated metadata features with static code attributes (e.g., import statements, function calls) in an ML model for PyPI, achieving higher accuracy than single-method approaches. Damodaran et al.~\cite{hybrid_damodaran_2022} compared hybrid methods against metadata and static analysis, finding that hybrid models reduce false positives by contextualizing code behavior. Afianian et al.~\cite{hybrid_afianian_2020} surveyed evasion techniques against static analysis, advocating for hybrid systems to counter advanced malware. Tools like MalOSS~\cite{maloss}, VirusTotal~\cite{VirusTotal}, and Packj~\cite{packj} use static code inspection along with repository metadata analysis, such as activity and contributor history. They assess the risk of malicious code in software packages. None of these tools specifically address or consider install-time execution threats, which have been increasingly observed in recent years~\cite{phpass2024, sharma2022pymafka}. To mitigate some of these limitations, hybrid analysis has been introduced, reducing false positive rates~\cite{hybrid_afianian_2020}. Nevertheless, challenges remain, including typosquatting, covert remote access activation, dynamic payload generation, and multiphase attack sequences~\cite{zheng2024robust}. As a result, researchers are increasingly turning to dynamic analysis to enhance detection capabilities and address these unresolved threats. A summary of existing tools for analyzing packages, along with their inputs and methods, is shown in Table~\ref{tab:existing_techniques}.

\begin{table}[h!]
\centering
\caption{Existing tools for analyzing PyPI packages.}
\begin{tabular}{|l|l|l|}
\hline
\textbf{Tool Name}                       & \textbf{Input}                & \textbf{Method}   \\ \hline
PyRadar~\cite{metadata_gao_2024}    & Package                       & Metadata             \\ \hline
Pyroma~\cite{metadata_pyroma}       & Package                       & Metadata             \\ \hline
Twine~\cite{metadata_twine}         & Package                       & Metadata             \\ \hline
Malware Checks~\cite{malwarechecks} & Setup Script File                 & Static               \\ \hline
OSSGadget~\cite{ossgadget}          & Package + Artifacts           & Static               \\ \hline
Bandit4Mal~\cite{bandit4mal}        & Package                       & Static               \\ \hline
VirusTotal~\cite{VirusTotal}        & Package + Artifacts           & Hybrid               \\ \hline
MalOSS~\cite{maloss}                & Package                       & Hybrid               \\ \hline
Packj~\cite{packj}                  & Package                       & Hybrid               \\ \hline
\textbf{DySec [Proposed]}                    & \textbf{Install-time Traces}           & \textbf{Dynamic}              \\ \hline
\end{tabular}
\label{tab:existing_techniques}
\end{table}

\textbf{eBPF-based analysis:} eBPF has emerged as a powerful tool for real-time system monitoring and is widely used across various domains to capture runtime and install-time behavior of applications~\cite{ebpf_bogle_2023, ebpf_eitani_2023}. For instance, it has been used for malware detection, network security, and performance tracing, enabling deeper visibility into system activities without significant overhead~\cite{ebpf_bogle_2023}. Higuchi and Kobayashi~\cite{evpf_Higuchi_2023} designed an eBPF-based ransomware detector using system call traces and ML, demonstrating low-latency detection. Similarly, Zhuravchak and Dudykevych~\cite{ebpf_Zhuravchak_2023} combined eBPF with NLP to classify ransomware behavior in real-time. Jia et al.~\cite{ebpf_jia_2023} proposed programmable security policies via eBPF, enabling customizable system call monitoring. Beyond threat detection, eBPF provides a novel approach to monitoring install-time behaviors, an area where traditional methods fall short. Eitani~\cite{ebpf_eitani_2023} introduced bpftrace, an eBPF-powered customization tool designed to identify install-time and runtime traces of applications. Unlike traditional tools such as Wireshark and Sysdig, eBPF enables lightweight install-time monitoring without requiring kernel modifications~\cite{ebpf_Zhuravchak_2023}. Additionally, its programmability in C or Python allows it to dynamically adapt to evolving security threats~\cite{ebpf_eitani_2023}.

However, traditional approaches relying on static code inspection or metadata analysis struggle to detect install-time attacks. These include typosquatting, covert remote access activation, dynamic payload generation, and multiphase attack sequences. The unique capabilities of eBPF highlight the need for frameworks that integrate eBPF with ML to analyze install-time traces for efficient malicious package detection.

\section{DySec Framework}
\label{Methodology}

\begin{figure}[htp!]
    \centering
    \includegraphics[width=0.85\linewidth]{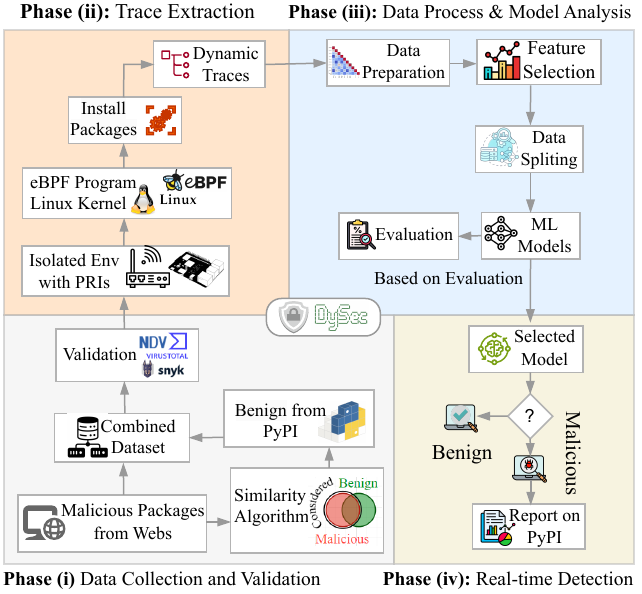}
    \caption{The overall workflow of DySec for detecting malicious packages.}
    \label{fig:The overall workflow}
\end{figure}

We propose DySec, a comprehensive evaluation framework for detecting malicious Python packages. It is a dynamic analysis framework that captures install-time behaviors to improve detection capabilities. Dysec comprised four key phases (i) dataset collection and validation, (ii) dynamic trace extraction, (iii) data processing and ML model evaluation, and (iv) detecting threats in real-time. The overall workflow of DySec is outlined in Figure~\ref{fig:The overall workflow}.

\textbf{Dataset collection and validation:} There is a scarcity of publicly available malicious PyPI package datasets, as such data is often scattered across multiple sources and not centrally maintained. This lack of structured datasets makes it challenging to obtain labeled samples for research and model training. Without a standardized dataset, researchers must rely on fragmented sources, increasing the risk of incomplete or misleading data. To address this challenge, we collected malicious packages from security blogs, threat intelligence feeds, and third-party repositories. However, not all sources reliably label threats, and some packages may be benign or mislabeled. This makes maliciousness validation a critical step to ensure the integrity of our dataset. To verify the collected samples, we cross-referenced them with 62 security vendors (e.g., Kaspersky, McAfee) via the existing VirusTotal Academic API~\cite{VirusTotal}. Additionally, we confirmed their absence from PyPI using its official API~\cite{pypi}, ensuring that our dataset contains genuinely malicious packages. This step enhances dataset reliability, eliminating potentially benign samples that could otherwise skew the analysis. Also, malicious packages often mimic benign ones by using similar names, descriptions, or metadata, making detection particularly challenging. 

Traditional detection methods relying solely on static analysis may struggle to distinguish between legitimate and maliciously altered versions of packages. To overcome this, we applied four similarity algorithms - String Matching (SM), Levenshtein Similarity (LS), Jaccard Similarity (JS), and Cosine Similarity (CS)~\cite{levenshtein1966binary}. These algorithms help identify benign counterparts on PyPI by measuring name and metadata similarities. To further validate this approach, the identified benign packages were verified as non-malicious using existing VirusTotal~\cite{VirusTotal}. This final step establishes a reliable baseline for comparative analysis, ensuring that the dataset accurately represents both malicious and benign samples.

\textbf{Dynamic trace extraction:} Monitoring the behavior of potentially malicious packages poses a significant challenge, as their execution could compromise system integrity. To mitigate this risk, an isolated Raspberry Pi cluster was deployed as a secure sandbox, ensuring reproducible and unbiased analysis of malicious Python packages while preventing any impact on production systems. This controlled environment enabled comprehensive trace monitoring to capture the dynamic behavior of installed packages. Each node in the cluster integrated eBPF into the Linux Kernel to capture six categories of dynamic traces during package installations, monitoring activity for 120 seconds both during and immediately after installation. Corrupted environments from installation failures were discarded, and affected packages were reinstalled in a fresh environment. Automated scripts were employed to aggregate and preprocess these traces into structured logs, optimizing data collection for further analysis. Algorithm~\ref{alg:dynamic_analysis} outlines the environment setup and trace extraction process for analyzing malicious Python packages in a distributed Raspberry Pi cluster. It automates secure and isolated package installation, install-time monitoring using eBPF tracing tools, and trace aggregation, ensuring efficient data collection for dynamic analysis.

\begin{algorithm}[htp!] \small
\SetAlgoLined
\DontPrintSemicolon
\LinesNumbered
\KwIn{Package, Pi $\{Pi_1,.,Pi_n\}$ with credentials}
\KwOut{Installation traces $\mathcal{T} = \{Trace_1,...,Trace_m\}$}
\textbf{Precondition}: All Pi run Python 3.8-3.12 isolated Env\;
\textbf{Step 1: Environment setup} \;
\Indp
1.1 Initialize and setup local virtual env: \texttt{venv} \;
1.2 Configure device: $\mathcal{D} \leftarrow [(host_i, user_i, key_i)]_{i=1}^n$\;
1.3 Load all the eBPF compiler tools\\
\Indm

\BlankLine
\textbf{Step 2: Distributed package deployment} \;
\ForEach{package $\pi_j \in \texttt{Package/}$}
{
        \If{$\pi_j \notin \{\texttt{.zip}, \texttt{.tar.gz}\}$}
    {
        \texttt{continue}\;
    }
    2.1 Extract $\pi_j$ metadata: $(name_j, version_j) \leftarrow \texttt{parse\_filename}(\pi_j)$ \;
    2.2 Select target device: $d_k \leftarrow \mathcal{D}[\texttt{rand}(|\mathcal{D}|)]$ \;
    2.3 Establish secure channel: \\
    \quad $session_{jk} \leftarrow \texttt{SSHConnect}(d_k.host, d_k.user, d_k.key)$\;
    2.4 Deploy package: \texttt{scp $\pi_j$ $\rightarrow$ $d_k$:/tmp/$name_j$}\;
}
\BlankLine
\textbf{Step 3: Installation traces} \;
\While{$\exists$ active $session_{jk}$}
{
    3.1 On $d_k$: Create \texttt{venv $name_j$ \&\& active $name_j$} \;
    3.2 Start tracing subsystems (120 sec): \\
    \quad filetop > /traces/$name_j$\_filetop.log 2>\&1 \& \\
    \quad opensnoop > /traces/$name_j$\_opens.log 2>\&1 \& \\
    \quad tcpconnect > /traces/$name_j$\_tcps.log 2>\&1 \& \\
    \quad syscall > /traces/$name_j$\_syscall.log 2>\&1 \& \\
    3.3 Execute installation:\\
    \quad pip install /$name_j$ $|$ tee /$name_j$\_inst.log\;
    3.4 Validate installation: \\
    \If{pip list $|$ grep $name_j$}
    {
        $status_j \leftarrow$ Success\;
    }
    \Else
    {
        $status_j \leftarrow$ Failed\;
        traces/$name_j$.log $\rightarrow$ /traces/$name_j$\_err.log\;
    }
    3.5 Terminate tracing: \\
    \quad kill filetop opensnoop tcpconnect syscall\;
}
\BlankLine
\textbf{Step 4: Trace aggregation} \;
\ForEach{$session_{jk}$}
{
    4.1 Copy traces:\\
    \quad scp $d_k$:/traces/$name_j$\ $\rightarrow$ Traces/$name_j$\;
    4.2 Annotate metadata:\\
    \quad $name_j$,$version_j$,$status_j$,$d_k$ $\rightarrow$ Traces/data.csv\;
    4.3 Clean remote env for another fresh install:\\
    \quad rm -rf /tmp/$name_j$ /opt/$name_j$\;
}
\textbf{Postcondition}: $\forall \pi_j \exists \hspace{5pt}Trace_j \in \mathcal{T}$ with: secure connections\\
\caption{Environment Setup and Trace Extraction}
\label{alg:dynamic_analysis}
\end{algorithm}

\textbf{Data processing and model evaluation:} The entire data processing and model evaluation pipeline was executed in an HPC environment to handle large-scale computations efficiently. The dataset, consisting of malicious and benign package traces, was partitioned into training, validation, and testing subsets to develop models for distinguishing malicious packages from benign ones at install time. It then underwent preprocessing to remove redundant features, optimize computational efficiency, and improve detection accuracy. The ML models were then trained on the processed data, with hyperparameters fine-tuned using the validation subset to mitigate overfitting. Finally, the best-performing model was selected based on performance metrics for real-world deployment.

\textbf{Detecting threats in real-time:} To assess robustness against emerging threats, a secondary analysis was conducted on packages updated or disclosed between July and November, extending beyond the initial dataset collected from November 2021 to June 2024. To ensure consistency, their dynamic traces were preprocessed in the same HPC environment using the same pipeline as the training data, maintaining uniform feature extraction and transformation. The best-performing model from the previous phase was deployed for classification, identifying benign and malicious packages. Any misclassified benign packages were flagged and reported to PyPI, ensuring transparency and proactive security enforcement.

\vspace{-5pt}
\section{DySec Design and Setup}
\label{Experimental Design and Setup}

This section provides an overview of the experimental setup, including the dataset used for analysis, and details the hyperparameters and training settings used during the study.

\vspace{-10pt}

\subsection{Environment Setup}

The experimental setup comprised of (i) an execution environment for package installation and trace collection, (ii) a network architecture for behavioral monitoring, and (iii)~an HPC cluster for large-scale ML analysis. The execution environment utilized 16 Raspberry Pi 4 devices, each running Ubuntu 24.4 LTS with Python 3.8–3.12 in virtualized environments. A private, isolated network with a dedicated router and switch was established to prevent external interference and control traffic flow. Behavioral monitoring was implemented using eBPF integrated into the Linux kernel (v6.8.0-1012-raspi), with tools such as bcc-tools, bpftool, and bpftrace for real-time tracing. For scalable ML analysis, an HPC cluster with 16-core CPUs, NVIDIA A100 GPUs, and 128 GB RAM was used to enable parallel processing and accelerate model training.

\vspace{-10pt}

\subsection{Dataset Overview}
This study analyzes 14,271 Python packages (7,127 malicious, 7,144 benign) to identify behavioral differences during installation. The dataset includes both successfully installed package traces (88\%) and those that exhibited installation failures or anomalies (12\%). This approach is critical, as even unsuccessful malicious installations—such as those causing system crashes, infinite loops, shutdowns, or unexpected authentication requests—pose operational threats and security risks. Additionally, some packages failed to generate metadata during installation, despite their dependencies being installed successfully. This raises security concerns, as installed dependencies could contain malicious components, allowing adversarial payloads to persist. In some cases, installing a required version inadvertently triggered the installation of a malicious package, further compromising system security.

To capture these variations, installation behaviors are categorized into normal, compatibility, and system-related behaviors, based on install-time characteristics observed in an isolated environment, as summarized in Table~\ref{table:characteristics}. 

\vspace{-5pt}

\begin{table}[htp!]
\caption{Characteristics of packages during install-time trace collection in an isolated environment using Python 3.8-3.12.}
\label{table:characteristics}
\centering
\begin{tabular}{|c|p{3.9cm}|c|c|}
\hline
\textbf{Type} & \textbf{Install-time Characteristics} & \textbf{Malicious} & \textbf{Benign} \\ 
\hline
\multirow{4}{*}{\rotatebox[origin=c]{90}{\textbf{Normal}}} 
& Successfully installed                  & 6,184  & 6,352  \\ 
& Unable to generate metadata             & 288    & 232    \\ 
& Missing setup files                     & 161    & 0      \\ 
& Failed build installable wheels         & 231    & 321    \\ 
\hline
\multirow{6}{*}{\rotatebox[origin=c]{90}{\textbf{Compatibility}}} 
& Mismatch distribution                   & 73     & 149    \\
& Requires a different version            & 121    & 53     \\ 
& Missing package version                 & 10     & 0      \\ 
& Unexpected auth request                 & 7      & 0      \\ 
& Missing installation module             & 23     & 0      \\ 
& Unicode file naming                     & 2      & 0      \\ 
\hline
\multirow{5}{*}{\rotatebox[origin=c]{90}{\textbf{System}}} 
& System freezing                         & 19     & 0      \\ 
& Infinity waiting                        & 4      & 0      \\ 
& System shutdown                         & 1      & 0      \\ 
& Version looping                         & 3      & 0      \\ 
& System prerequisites required           & 0      & 37     \\ 
\hline
\multicolumn{2}{|r|}{\textbf{Total}}      & \textbf{7,127}  & \textbf{7,144}   \\
\hline
\end{tabular}
\end{table}

To comprehensively analyze these installation behaviors, eBPF-based monitoring was employed to capture real-time system interactions and detect anomalies. eBPF provides various probes for security and monitoring, enabling the detection of malicious execution patterns, unauthorized network access, and system manipulation~\cite{ebpf_eitani_2023}. For instance, file system monitoring probes include \textit{opensnoop} and \textit{filetop}, while network activity can be monitored using \textit{tcptop} and \textit{tcpconnect}. Additionally, system call monitoring utilizes probes like \textit{syscount} and \textit{syscall}, whereas memory monitoring includes \textit{memleak} and \textit{biolatency}, among others~\cite{ebpf_eitani_2023, ebpf_bogle_2023}.

To enhance malware detection, these eBPF features have been widely utilized in scenarios such as network intrusion detection and ransomware analysis~\cite{ebpf_bogle_2023, evpf_Higuchi_2023, ebpf_Zhuravchak_2023}. However, metadata and static analysis methods have limitations in detecting sophisticated threats due to their inability to capture install-time behaviors. To address this critical gap, user and kernel-level probes were selected to analyze system activity from multiple perspectives, ensuring a comprehensive behavioral analysis for enhanced threat detection. Each of these probes is designed to capture specific aspects of execution, allowing the identification of malicious patterns that traditional static approaches may overlook. The selected eBPF probe categories for detecting malicious packages are as follows:

\begin{itemize}
    \item \textbf{FiletopTraces}: Monitored file read/write processes (e.g., Read\_Processes, Write\_Data\_Transfer), directly linking to missing critical files like setup.py or pyproject.toml.
    \item \textbf{InstallTraces}: Tracked dependency chains (direct/indirect), exposing dependency mismatches (e.g., `No match distribution') used in dependency confusion attacks.
    \item \textbf{OpensnoopTraces}: Logged directory access patterns (e.g., /root, /usr, /tmp, /sys), revealing unauthorized access to restricted paths like /root/.ssh.
    \item \textbf{TCPTraces}: Captured network interactions (remote IP/port access, connection states), identifying anomalous ports (e.g., port 6667) tied to malicious payload delivery.
    \item \textbf{SystemCallTraces}: Recorded OS-level operations (file I/O, network, security, process management), correlating with system-level sabotage (e.g., shutdowns, freezes).
    \item \textbf{PatternTraces}: Aggregated sequential behaviors (e.g., Pattern\_4: network socket creation, Pattern\_5: process creation), detecting infinite loops or version cycling.
\end{itemize}

A detailed breakdown of these features, along with dataset examples, is provided in Appendix A1, offering further insight into how these traces contribute to malicious detection.

\subsubsection{Data Preparation}

The install-time traces underwent a preprocessing pipeline that included data cleaning, integration, encoding, and transformation to ensure consistency and compatibility with ML models for accurate classification.

\textbf{Data cleaning and integration:} To ensure dataset integrity, systematic data cleaning and integration were performed, focusing on three key steps: removing duplicate entries, filtering incomplete installation traces, and standardizing paths for feature encoding. Duplicate entries, such as typosquatted malicious package variants (e.g., reverse-shell and reverse\_shell), were eliminated to prevent redundancy. Incomplete installation traces caused by Raspberry Pi cluster power issues and internet disruptions were filtered out to maintain dataset reliability. Additionally, all package installations were executed across 16 Raspberry Pi devices using a homogeneous path structure, ensuring consistency in feature encoding while avoiding bias from device-specific identifiers.

\textbf{Data encoding and transformation:} The extracted features were categorized into two types - categorical and numerical. Categorical features included system call sequences, which serve as fundamental indicators of program behavior, capturing interactions between package installations and the operating system. These sequences provide valuable insights into behavioral traits that aid in detection. For example, system call sequences such as \textit{newfstatat->openat->fstat} and \textit{socket->bind->listen} frequently appear in specific operational contexts. To capture these sequential behaviors, n-grams were employed to represent contiguous sequences of n events. This approach effectively models execution patterns, enabling differentiation between benign and malicious behaviors. The system call sequences were then vectorized into frequency-based representations, allowing ML models to recognize behavioral patterns. By preserving contextual meanings, the n-grams technique enhances the ability of ML models to learn and distinguish system behaviors~\cite{Masud2016}.

\begin{figure}[htp!]
    \centering
    \includegraphics[width=0.9\linewidth]{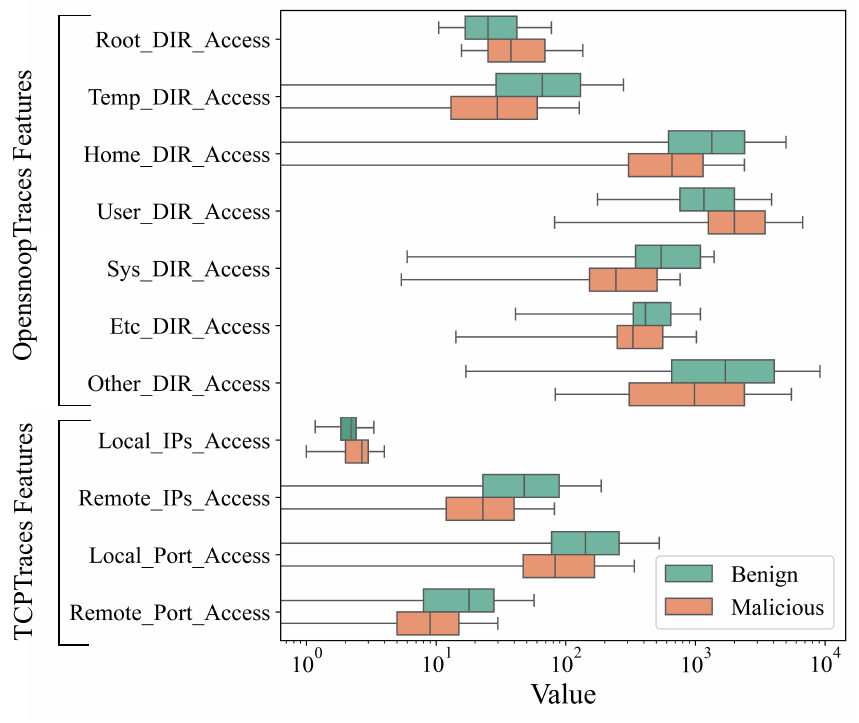}
    \caption{A boxplot of numerical features from OpensnoopTraces and TCPTraces (before rescaling) on a logarithmic x-axis shows a significant median difference between benign and malicious packages.
    }
    \label{fig:Boxplot}
\end{figure}

Numerical features, such as \textit{Root\_DIR\_Access} and \textit{Remote\_Port\_ Access}, exhibited significant scale variations, spanning $10^0$ to $10^4$ (Figure~\ref{fig:Boxplot}). Such disparities can bias ML models, as algorithms like RF and SVM often prioritize features with larger magnitudes, potentially leading to skewed performance. To mitigate this, \textit{Minimum–Maximum Normalization} method ~\cite{hastie2009elements} was applied, ensuring all numerical features were scaled to a comparable range while preserving their intrinsic distributions.

\vspace{-15pt}

\subsection{Features Extraction and Selection} 
The initial set consisted of 62 Candidate Features (CFs), capturing various installation behaviors across six trace categories. However, 22 features were identified as Dependent Features (DFs) due to high correlation (\(|r| > 0.50\)) and were removed using \textit{Pearson Correlation Analysis (r)}~\cite{pearson1895} to mitigate multicollinearity. This filtering step resulted in 40 Independent Features (IDFs) for further evaluation. To illustrate, in FiletopTraces, five features were retained, while four were removed due to high correlation. In contrast, PatternTraces exhibited no strong correlations (\(|r| \leq 0.49\)). This highlights the unique behaviors, reinforcing their importance in detecting malicious patterns.

To further refine the feature set, Importance Scores (IMS) from four ML models were used to select the most relevant features~\cite{hastie2009elements, Mehedi2023DependableApproach}. Features with an IMS \(> 0.05\) in at least one model were retained to ensure flexible feature selection. The final SEF set was determined using Equation~\ref{sef}.

\vspace{-10pt}

\begin{align} \small
    SEF = & \Bigg\{ f \in CF \mid |r_f| < 0.50, \max_{m \in M} IMS_m(f) > 0.05 \Bigg\} \cap \notag \\
    & \Bigg\{ f \mid \max_{m \in M} IMS_m(f) > 0.08 \Bigg\}
    \label{sef}
\end{align}
\vspace{-5pt}
\noindent where:
\begin{itemize}
    \item \( CF \) represents the initial set of candidate features.
    \item \( |r_f| < 0.50 \) removes highly correlated features.
    \item \( IMS_m(f) \) is the importance score of feature \( f \) evaluated by model \( m \), where \( M \) is the set of all ML models used.
    \item \( \max_{m \in M} IMS_m(f) > 0.05 \) retains a feature if at least one model assigns it an importance score > 0.05.
    \item \( \max_{m \in M} IMS_m(f) > 0.08 \) further filters the features, selecting only those with a higher importance threshold.
\end{itemize}

Table~\ref{table:traces} provides an overview of the feature selection process. For example, all 10 PatternTraces features surpassed the IMS baseline threshold \(> 0.05\) and met the Selected Engineered Features (SEF) threshold \(> 0.08\). In contrast, SystemCallTraces retained only 6 out of 17 features, as one feature failed to meet the SEF threshold in RF, leading to its exclusion. 
After filtering out correlated and low-importance features, the CombinedTraces feature set was reduced from 62 CFs to 36 SEFs, a 58\% reduction. 
The final feature set prioritized low inter-feature correlation, high-importance scores across models, and computational efficiency.

\vspace{-8pt}

\begin{table}[h!]
\caption{Feature selection and reduction process, filtering correlated features and selecting important features.}
\label{table:traces}
\centering
\begin{tabular}{|l|c|c|c|c|c|c|c|l|}
\hline
\multirow{2}{*}{\textbf{Traces Category}} & 
{\rotatebox{90}{\multirow{2}{*}{\textbf{CF (Raw Data)}}}} & 
{\rotatebox{90}{\multirow{2}{*}{\textbf{DF (|r| > 0.50)}}}} & 
{\rotatebox{90}{\multirow{2}{*}{\textbf{IDF (|r| < 0.50)}}}} & 
\multicolumn{4}{c|} 
{\makecell{\textbf{IMF (IMS > 0.05)}}} & 
{\rotatebox{90}{\multirow{2}{*}{\textbf{SEF (IMS > 0.08) }}}}
\\ 
\cline{5-8}
&  &  &  & \textbf{RF} & \textbf{DT} & \textbf{SVM} & \textbf{GB} & \\ 
\hline
FiletopTraces      & 9 & 4 & 5 & 5 & 4  & 3  & 5 & 5 \\ 
\hline
InstallTraces      & 6 & 3 & 3 & 3  & 2  & 2  & 2 & 3 \\ 
\hline
OpensnoopTraces    & 12 & 5 & 7 & 7 & 6  & 5  & 7 & 7 \\  
\hline
TCPTraces          & 8 & 2 & 6 & 6  & 6  & 4  & 4 & 5 \\ 
\hline
SystemCallTraces  & 17 & 6 & 11 & 7  & 6  & 5  & 6 & 6 \\  
\hline
PatternTraces      & 10 & 0 & 10 & 10 & 9 & 7 & 9 & 10 \\ 
\hline
\textbf{CombinedTraces}  & \textbf{62} & \textbf{22} & \textbf{40} & \textbf{37} & \textbf{35}  & \textbf{28}  & \textbf{33} & \textbf{36} \\ 
\hline
\end{tabular}
\begin{flushleft}
\footnotesize Note: Candidate Features (CF); Dependent Features (DF); Independent Features (IDF); Important Features (IMF); Selected Features (SEF).
\end{flushleft}
\end{table}

\vspace{-10pt}

\subsection{Learning Package Maliciousness}

To evaluate the effectiveness of ML models in detecting malicious packages, we assessed four models - Random Forest (RF), Decision Tree (DT), Support Vector Machine (SVM), and Gradient Boosting (GB). Each model was configured with hyperparameters that balance simplicity and performance. The RF model was configured with 100 trees and a maximum depth of 8, ensuring a trade-off between overfitting mitigation and computational efficiency. For DT, a moderate depth (max\_depth=8) and conservative splitting criteria (min\_samples\_split=10) were applied to enhance generalizability. SVM was optimized with a linear kernel, facilitating efficient training and probabilistic outputs. GB utilized 100 weak learners with a shallow tree depth (max\_depth=5) and a learning rate of 0.1 to refine sequential error correction.

The dataset was partitioned into 70\% training, 15\% validation, and 15\% test sets. To improve generalizability, we used 5-fold stratified cross-validation to preserve class distribution across splits. The validation set was used for hyperparameter tuning (e.g., optimizing RF’s n\_estimators or SVM’s kernel), while the test set evaluated the final performance.

\vspace{-5pt}
\section{Evaluation}
\label{Evaluation}

This section evaluates the overall performance of the selected models through a comprehensive analysis.

\vspace{-8pt}

\subsection{Evaluation Metrics}

The evaluation of the selected models used four key metrics - accuracy, precision, recall, and F1-score - to assess performance across various scenarios. Accuracy measured overall correctness, while precision minimized false positives (FP), where benign packages were incorrectly classified as malicious. Recall reduced false negatives (FN), where truly malicious packages were incorrectly classified as benign. The F1-score balanced precision and recall, addressing class imbalance to provide a comprehensive assessment of model robustness and practical applicability. The evaluation was guided by the following Research Questions (RQs): 

\begin{itemize} 
    \item \textbf{RQ1}: What are the specific feature sets that DySec uses for malicious package detection?
    \item \textbf{RQ2}: Which patterns distinguish malicious packages from benign packages?
    \item \textbf{RQ3}: Is DySec accurate enough to be practical? 
\end{itemize}
\vspace{-10pt}
\subsection{RQ1: What are the specific feature sets that DySec uses for malicious package detection?}

The performance evaluation of ML models across diverse feature sets presents important insights into how trace types impact malicious package detection. This section analyzes these variations, highlighting which traces enhance or degrade performance and why certain models perform better with specific trace types.

\subsubsection{Feature set performance analysis}

Feature set performance analysis evaluates the effectiveness of selected features in a model, measuring their impact on overall predictive power. It helps in feature selection to enhance model performance.

\textbf{Combined traces analysis:} The integration of six trace types - FiletopTraces, InstallTraces, OpensnoopTraces, TCPTraces, SystemCallTraces, and PatternTraces - into CombinedTraces consistently achieves the highest performance across all models, as shown in Table~\ref{tab:Performance evaluation results}. For instance, RF attains 95.99\% accuracy and 96.00\% precision with CombinedTraces, surpassing all standalone traces. This improvement is due to the complementary nature of different traces, where each contributes a distinct behavioral dimension-resource monitoring, network activity, or system-level anomalies.
    
In detail, FiletopTraces captures resource usage trends (e.g., file read/write frequency), while OpensnoopTraces detects unauthorized directory exploration through access patterns. TCPTraces provide insights into network behavior by tracking suspicious IP or port connections, often linked to data exfiltration or command-and-control activity. However, these traces alone may not capture deeper system interactions. SystemCallTraces bridges this gap by recording OS-level activities, such as privilege escalation attempts. Meanwhile, PatternTraces aggregate sequential anomalies - such as rapid file modifications followed by network requests - revealing multi-stage attack strategies. By integrating these diverse behavioral signals, CombinedTraces provides a holistic view, significantly improving classification accuracy over individual traces. While InstallTraces contributes less to detection accuracy due to metadata overlap between benign and malicious packages, their inclusion broadens trace coverage, minimizing classification blind spots.
    
\textbf{Standalone trace analysis:} The performance of standalone traces varies significantly, affecting their effectiveness in malicious package detection. PatternTraces achieved the highest standalone performance (RF accuracy 94.62\%, and F1 Score 94.61\%), demonstrating their ability to capture aggregated high-level behaviors such as directory and network access sequences that distinguish malicious packages. SystemCallTraces, while effective (RF accuracy 88.51\%), focuses on low-level OS interactions that may require additional context for optimal classification. FiletopTraces (RF accuracy 92.01\%) and TCP Traces (83.74\%) provided moderate performance by capturing resource usage and network anomalies, though their standalone effectiveness remained limited. InstallTraces is the weakest performer (RF accuracy 69.45\%), likely due to significant overlap in installation dependencies and metadata between benign and malicious packages, reducing their discriminative power.

\begin{table*}[htp!]
\caption{Performance evaluation results for ML algorithms on different features; bold indicates the overall highest values, upward arrow (↑) the second-best, downward arrow (↓) the third-best, and underlined the lowest values.}
\label{tab:Performance evaluation results}
\centering
\begin{tabularx}{\textwidth}{|c|l|>{\centering\arraybackslash}X|>{\centering\arraybackslash}X|>{\centering\arraybackslash}X|>{\centering\arraybackslash}X|>{\centering\arraybackslash}X|>{\centering\arraybackslash}X|>{\centering\arraybackslash}X|}
\hline
\textbf{ML} & \textbf{Metrics} & \textbf{Filetop Traces} & \textbf{Install Traces} & \textbf{Opensnoop Traces} & \textbf{TCP Traces} & \textbf{System Call Traces} & \textbf{Pattern Traces} & \textbf{Combined Traces} \\ 
\hline
\multirow{4}{*}{\rotatebox{90}{RF}}
& Accuracy        & 0.9201 & 0.6945 & 0.9355 & 0.8374 & 0.8851 & 0.9462 & \textbf{0.9599} \\ 
& Precision       & 0.9210 & 0.8028 & 0.9362 & 0.8374 & 0.8851 & 0.9495 & \textbf{0.9600} \\ 
& Recall          & 0.9201 & 0.6945 & 0.9355 & 0.8374 & 0.8851 & 0.9462 & \textbf{0.9599} \\ 
& F1 Score        & 0.9200 & 0.6647 & 0.9355 & 0.8374 & 0.8851 & 0.9461 & \textbf{0.9602} \\ 
\hline
\multirow{4}{*}{\rotatebox{90}{DT}} 
& Accuracy        & 0.8687 & 0.6950 & 0.9135 & 0.8113 & 0.8841 & 0.9462 ↓ & 0.9402 \\ 
& Precision       & 0.8687 & 0.8084 & 0.9136 & 0.8121 & 0.8841 & 0.9495 ↓ & 0.9436 \\ 
& Recall          & 0.8687 & 0.6950 & 0.9135 & 0.8113 & 0.8841 & 0.9462 ↓ & 0.9402 \\ 
& F1 Score        & 0.8687 & 0.6643 & 0.9135 & 0.8111 & 0.8841 & 0.9461 ↓ & 0.9428 \\ 
\hline
\multirow{4}{*}{\rotatebox{90}{SVM}} 
& Accuracy        & 0.8977 & 0.6865 & 0.8005 & 0.8047 & 0.8556 & 0.9453 & 0.9528 ↑ \\ 
& Precision       & 0.8985 & 0.8065 & 0.8165 & 0.8055 & 0.8557 & 0.9487 & 0.9530 ↑ \\ 
& Recall          & 0.8977 & 0.6865 & 0.8005 & 0.8047 & 0.8556 & 0.9453 & 0.9528 ↑ \\ 
& F1 Score        & 0.8976 & 0.6539 & 0.7979 & 0.8046 & 0.8556 & 0.9452 & 0.9523 ↑ \\ 
\hline
\multirow{4}{*}{\rotatebox{90}{GB}}
& Accuracy        & 0.8738 & 0.6716 & 0.9154 & 0.8047 & 0.8561 & \underline{0.9458} & 0.9411 \\ 
& Precision       & 0.8742 & 0.7931 & 0.9167 & 0.8072 & 0.8562 & \underline{0.9488} & 0.9442 \\ 
& Recall          & 0.8738 & 0.6716 & 0.9154 & 0.8047 & 0.8561 & 0.9458 & \underline{0.9461} \\ 
& F1 Score        & 0.8738 & 0.6339 & 0.9153 & 0.8043 & 0.8561 & \underline{0.9457} & 0.9435 \\ 
\hline
\end{tabularx}
\end{table*}

\subsubsection{Algorithm-based performance analysis}

\begin{figure}[htp!]
    \centering
    \includegraphics[width=.92\linewidth]{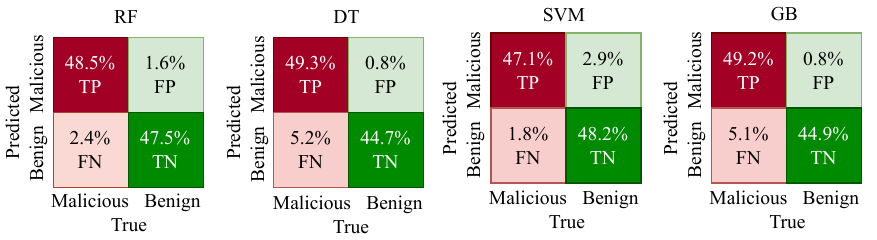}
    \caption{Confusion matrix of selected ML models on CombinedTraces.}
    \label{fig:ConfusionMatrix}
\end{figure}

Among the evaluated algorithms RF, DT, SVM, and GB, the RF-based classifier demonstrates superior accuracy with PatternTraces, achieving 94.62\% by utlizing its ensemble structure to detect intricate behavioral patterns effectively. Table~\ref{tab:Performance evaluation results} compares the performance metrics across all classifiers, highlighting RF’s consistent superiority in accuracy, precision, recall, and F1-score across most traces. This advantage stems from its ability to handle high-dimensional data, mitigate overfitting through feature randomness, and aggregate diverse DTs. For example, RF’s precision with CombinedTraces (96.00\%) surpasses DT (94.36\%), SVM (95.30\%), and GB (94.42\%). Its lower FP rate of 1.6\% and FN rate of 2.4\%, as shown in the confusion matrix (Figure~\ref{fig:ConfusionMatrix}), reinforce its reliability in minimizing classification errors and accurately detecting malicious packages.

\vspace{-5pt}

\vspace{-5pt}

Unlike RF, DT shows signs of overfitting, as evidenced by its strong performance with PatternTraces at 94.62\% but a slight drop to 94.02\% with CombinedTraces. This suggests that DT relies more on dataset-specific patterns rather than generalizable features, leading to reduced adaptability when presented with a more diverse dataset. Finally, SVM achieves strong accuracy with CombinedTraces (95.28\%) by utilizing a linear kernel, effectively distinguishing classes in high-dimensional space through optimized decision boundaries. However, its computational cost (4.4151s test time in Table~\ref{tab:Performance comparison of metadat}) makes it less practical for large-scale detection. GB, while effective with PatternTraces (94.58\%) due to its iterative error correction, struggles with SystemCallTraces (85.61\%), indicating sensitivity to trace-specific noise.

\begin{figure}[htp!]
    \centering
    \includegraphics[width=0.92\linewidth]{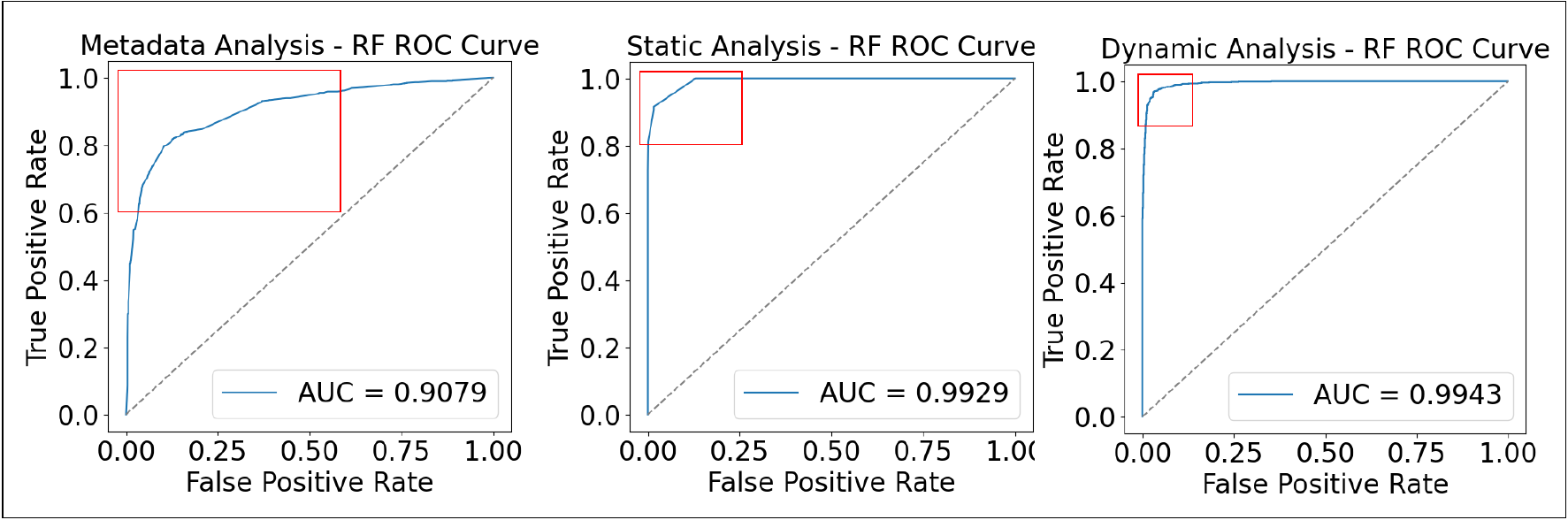}
    \caption{ROC AUC curve comparing metadata, static, and dynamic analysis using the best-performing RF model.}
    \label{fig:ROC AUC curve}
\end{figure}
\vspace{-5pt}

\subsubsection{Comparison with state-of-the-art methods}

Malicious package detection requires a balance between accuracy, speed, and generalization. While ML-based metadata and static methods have been widely studied, they struggle to detect sophisticated threats due to their limited behavioral insights. This section compares DySec with the ML-based metadata method proposed by Halder et al.~\cite{metadata_halder_2024} and the ML-based static method introduced by Samaana et al.~\cite{hybrid_samaana_2024}, highlighting key performance trade-offs and DySec’s advantages in real-world scenarios. To ensure a fair comparison with state-of-the-art methods, the same model, feature selection techniques, and hyperparameters from the original research were applied to the collected metadata and static dataset.

As shown in Table~\ref{tab:Performance comparison of metadat}, DySec with CombinedTraces and RF outperformed metadata and static methods, achieving 95.99\% accuracy, 96.02\% F1 Score, and 99.37\% ROC AUC (Figure~\ref{fig:ROC AUC curve}). Its strength lies in analyzing install-time behavior, enabling the detection of subtle anomalies. This resulted in a high TP rate (48.5\%) and TN rate (47.5\%), while minimizing the FP rate (1.6\%) and FN rate (2.4\%), as shown in Figure~\ref{fig:ConfusionMatrix}. These results confirm DySec’s effectiveness in detecting malicious packages while reducing misclassifications.

\begin{table*}[htp!]
\centering
\caption{Performance comparison of metadata analysis (proposed by Halder et al.~\cite{metadata_halder_2024}), static analysis (proposed by Samaana et al.~\cite{hybrid_samaana_2024}), and dynamic analysis (DySec) using popular ML methods; bold indicates the overall best values.}
\label{tab:Performance comparison of metadat}
\centering
\begin{tabularx}{1.0 \textwidth}{|c|>{\centering\arraybackslash}c|>{\centering\arraybackslash}c|>{\centering\arraybackslash}X|>{\centering\arraybackslash}X|>{\centering\arraybackslash}X|>{\centering\arraybackslash}X|>{\centering\arraybackslash}X|c|c|c|c|}
\hline
\multirow{2}{*} {\textbf{Methods}} & \multirow{2}{*}{\textbf{Data Source}} & \multirow{2}{*}{\textbf{Features}} & \multirow{2}{*}\makecell{\textbf{ML Model}} & \multirow{2}{*}\makecell{\textbf{Test Accu.}} & \multirow{2}{*}\makecell{\textbf{F1 \hspace{5pt} Score}} & \multirow{2}{*}\makecell {\textbf{Test Time (s)}} & \multicolumn{4}{c|} {\textbf{Confusion Matrix}} \\ \cline{8-11}
 &  &  &  & & &  & \textbf{TP} ↑ & \textbf{TN} ↑& \textbf{FP} ↓& \textbf{FN} ↓\\ 
\hline
\multirow{4}{*}{\makecell{Metadata \\Analysis~\cite{metadata_halder_2024})}}
 & \multirow{4}{*}{\makecell{File \\Properties}} & \multirow{4}{*}{57} & RF & 0.8444 & 0.8481 & 0.4210 & 930 & 878 & 142 & 191 \\
 &  &  & DT & 0.8393 & 0.8436 & 0.0849 & 928 & 869 & 144 & 200 \\
 &  &  & SVM & 0.8047 & 0.8160 & 4.6816 & 927 & 796 & 145 & 273 \\
 &  &  & GB & 0.8346 & 0.8425 & 0.0780 & 947 & 840 & 125 & 229 \\ 
\hline
\multirow{4}{*}{\makecell{Static \\Analysis~\cite{hybrid_samaana_2024}}}
 & \multirow{4}{*}{\makecell{File and Source \\or Binary Code}} & \multirow{4}{*}{21} & RF & 0.9514 & 0.9524 & 0.3676 & 1042 & 995 & 30 & 74 \\
 &  &  & DT & 0.9514 & 0.9529 & 0.0410 & 1054 & 983 & 18 & 86 \\
 &  &  & SVM & 0.9532 & 0.9530 & 3.0306 & 1014 & 1027 & 58 & 42 \\
 &  &  & GB & 0.9490 & 0.9508 & \textbf{0.0260} & 1054 & 978 & 18 & 91 \\
\hline
\multirow{4}{*}{\makecell{Dynamic \\ Analysis [DySec]}}
& \multirow{4}{*}{\makecell{Install-time \\ Behavior}} & \multirow{4}{*}{36} & RF & \textbf{0.9599} & \textbf{0.9602} & 0.4183 & 1038 & 1017 & 34 & 52\\
 &  &  & DT & 0.9402 & 0.9428 & 0.0588 & \textbf{1055} & 958 & \textbf{17} & 111 \\
 &  &  & SVM & 0.9528 & 0.9523 & 4.4151 & 1009 & \textbf{1031} & 63 & \textbf{38} \\
 &  &  & GB & 0.9411 & 0.9435 & 0.0355 & 1054 & 961 & 18 & 108 \\ 
\hline
\end{tabularx}
\end{table*}

\vspace{-2pt}

In contrast, metadata analysis, based on the method proposed by Halder et al.~\cite{metadata_halder_2024}, suffered from a high FP rate of 142 (6.6\%) due to its reliance on superficial file properties, leading to frequent misclassifications. Similarly, static analysis, following the approach of Samaana et al.~\cite{hybrid_samaana_2024}, lacked install-time behavioral context, resulting in a higher FN rate of 74 (3.5\%). The absence of behavioral insights weakened its ability to differentiate between malicious and benign packages, reducing overall detection accuracy.

Beyond accuracy, an effective security solution must also balance speed and scalability. DySec’s RF model achieved a test time of 0.4183s, providing high accuracy with computational efficiency. While metadata analysis was the fastest, it suffered from high FP and FN rates, making it impractical for large-scale deployments. Similarly, static analysis had a slightly faster test time (0.3676s) but failed to detect install-time anomalies, leading to higher FN rates. DySec’s ability to balance high accuracy, low FP/FN rates, and reasonable test times makes it well-suited for large-scale ecosystems like PyPI, where real-time detection and scalability are critical.

While speed and scalability are essential, a robust security system must also generalize well to new and evolving threats. A key strength of DySec is its ability to detect previously unseen malware by analyzing install-time execution behavior. Unlike metadata and static analysis, which rely on predefined signatures and patterns, DySec identifies suspicious execution patterns that deviate from expected behavior. As shown in Figure~\ref{fig:ROC AUC curve}, DySec achieved the highest ROC AUC, demonstrating strong discriminatory power in detecting malicious packages. While metadata and static methods performed well on known threats, they struggled with adversarial samples, limiting their effectiveness. By leveraging install-time behavior, DySec improves its ability to detect emerging attack variations, making it a more resilient solution against sophisticated cyber threats.

\vspace{-2pt}

\subsection{RQ2: Which patterns distinguish malicious packages from benign packages?}

Distinguishing between benign and malicious system call patterns is challenging, as some sequences appear in both contexts. The same pattern may indicate routine activity in one case and adversarial intent in another, depending on its execution context. To address this, DySec identifies unique and independent system call patterns and evaluates their contributions to classification.

\begin{figure}[htp!]
    \centering
    \includegraphics[width=.92\linewidth]{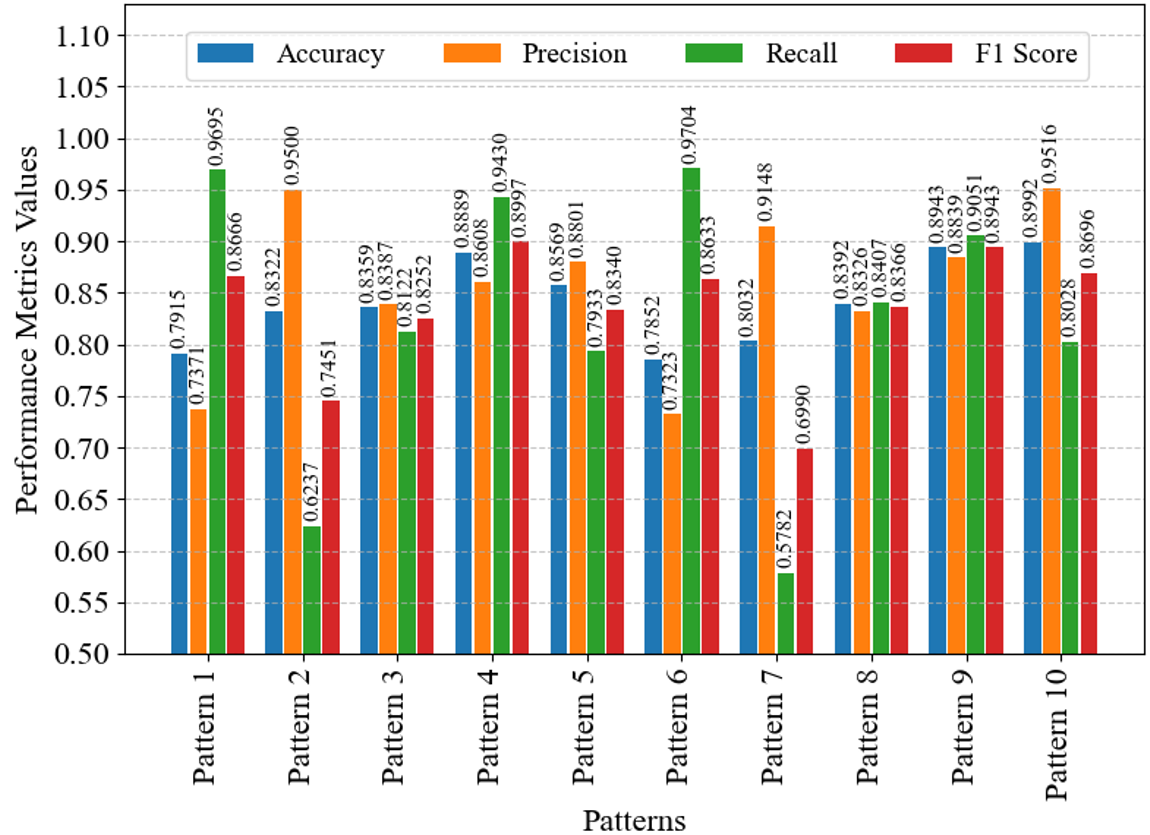}
    \caption{Comparison metrics across SystemCall patterns using RF model.}
    \label{fig:Comparison metrics across SystemCallTraces}
\end{figure}

To validate its effectiveness, the ten PatternTraces categories were analyzed to assess their role in differentiating benign and malicious packages. Figure~\ref{fig:Comparison metrics across SystemCallTraces} presents a comparative analysis of these categories, measuring their effectiveness in classification using key performance metrics. The results indicate that individual pattern categories contribute uniquely to the detection process, reinforcing their role in identifying malicious behaviors.

Each of these categories consists of multiple system call patterns that represent specific system behaviors. During the evaluation of PatternTraces, 56 distinct behavioral patterns were identified, each mapped to one of ten predefined categories. These patterns play a crucial role in classification by capturing fundamental differences in system behavior between benign and malicious packages. To ensure these patterns captured unique behavioral characteristics rather than redundant information, independence was assessed using correlation analysis. The most influential system call patterns were then identified by computing feature importance scores using the RF classifier’s \textit{feature\_importances\_} attribute, which assigns importance based on each feature’s impact on predictive performance~\cite{scikit-learn-feature-importance}. \textit{feature\_importances\_} was chosen over \textit{permutation\_importance} because PatternTraces columns contain only 56 unique patterns, indicating low cardinality. While both methods are applicable, \textit{feature\_importances\_} is computationally more efficient for this scenario and aligns well with the tree-based structure of the RF classifier~\cite{scikit-learn-feature-importance}.

\begin{table*}[ht]
\centering
\caption{Performance metrics for the top ten system call patterns.}
\begin{tabular}{|p{5cm}|p{6cm}|c|c|c|c|}
\hline
\textbf{Pattern Type and Specific Pattern} & \textbf{Pattern} & \textbf{Accuracy} & \textbf{Precision} & \textbf{Recall} & \textbf{F1 Score} \\ \hline
Pattern\_10: Error Handling & \textit{newfstatat -> openat -> fstat -> error=ENOENT} & 0.9954 & 0.9954 & 0.9954 & 0.9977 \\ \hline
Pattern\_1: File Metadata Retrieval & \textit{newfstatat -> openat -> fstat -> lseek -> ioctl} & 0.9844 & 0.9914 & 0.9922 & 0.9866 \\ \hline
Pattern\_2: Reading data from file & \textit{read -> read -> read -> newfstatat} & 0.9730 & 0.9875 & 0.9863 & 0.9863 \\ \hline
Pattern\_4: Network Socket Creation & \textit{socket -> bind -> listen -> accept -> execve}  & 0.9690 & 0.9835 & 0.9843 & 0.9843 \\ \hline
Pattern\_6: Memory Mapping & \textit{openat -> mmap -> ioctl -> prctl -> no-fd} & 0.9679 & 0.9796 & 0.9837 & 0.9837 \\ \hline
Pattern\_7: File Descriptor Management & \textit{fcntl -> fcntl -> close -> no-error -> fd=1} & 0.9634 & 0.9757 & 0.9814 & 0.9814 \\ \hline
Pattern\_1: File Metadata Retrieval & \textit{openat -> fstat -> ioctl} & 0.9577 & 0.9717 & 0.9784 & 0.9784 \\ \hline
Pattern\_1: File Metadata Retrieval & \textit{close -> newfstatat -> no-fd} & 0.9286 & 0.9441 & 0.9630 & 0.9630 \\ \hline
Pattern\_5: Creating a New Process & \textit{ioctl -> setresuid -> setresgid -> execve} & 0.9294 & 0.9480 & 0.9634 & 0.9634 \\ \hline
Pattern\_9: File Locking & \textit{openat -> fstat -> fcntl -> no-fd} & 0.9259 & 0.9401 & 0.9615 & 0.9615 \\ \hline
\end{tabular}
\label{tab:pattern_performance}
\end{table*}

Table~\ref{tab:pattern_performance} presents the top ten system call patterns observed across benign and malicious packages. While the table does not explicitly label each pattern as benign or malicious, their system call patterns provide meaningful insights. By analyzing their occurrence, frequency, and sequence, these patterns serve as behavioral signatures, distinguishing system calls that are more prevalent in malicious activities from those typical of benign operations. A detailed analysis follows, in which each pattern is examined in depth to determine whether it is associated with benign or malicious behavior based on its behavioral signatures and classification impact. Key distinctions in behavioral signatures:

\textbf{1. Error handling:} Error patterns help distinguish between malicious probing and legitimate file access. Our analysis identifies frequent errors, such as error=ENOENT, as strong signals of \textbf{malicious} behavior. For instance, Pattern\_10 (\textit{newfstatat->openat->fstat->error=ENOENT}) achieves an F1 score of 0.9977. In contrast, our analysis identifies error-free such as Pattern\_2: (\textit{read->read->read->newfstatat}) as a characteristic of routine file reads in \textbf{legitimate} operations. Their high recall and precision indicate their reliability in distinguishing normal activity from malicious probing.

\textbf{2. Metadata and read/write operation:} Pattern\_7 (\textit{fcntl->fcntl->close->no-error->fd=1}) indicates \textbf{malicious} behavior, suggesting file descriptor manipulation for privilege escalation, with a high F1 score of 0.9814. Conversely, Pattern\_1 (\textit{newfstatat->openat->fstat->lseek->ioctl}) reflects \textbf{benign} operations, showing strong accuracy (0.9844) and precision (0.9914), typical of safe installation workflows.

\textbf{3. Network and file behaviors:} Pattern\_4 (\textit{socket->bind->listen->accept->execve}) highlights \textbf{malicious} behavior, indicating potential command-and-control (C2) communication, with a strong F1 score of 0.9843. Similarly, Pattern\_9 (\textit{openat->fstat->fcntl->no-fd}) signals ransomware activity, particularly through file locking techniques. In contrast, Pattern\_6 (\textit{openat->mmap->ioctl->prctl->no-fd}) represents \textbf{benign} operations, reflecting memory mapping for dependency loading. Its high precision (0.9796) and recall (0.9837) demonstrate its effectiveness in identifying normal processes and avoiding FP.

\textbf{4. Structural independence:} The clear separation and minimal overlap between \textbf{benign} and malicious patterns enhance classification accuracy. For example, Pattern\_2 (\textit{read->read->read->newfstatat}) remains distinct from \textbf{malicious} probing behaviors like Pattern\_10 (\textit{error=ENOENT}), as reflected in their contrasting performance metrics.
 
\textbf{5. Robust classification:} The independence of high-performing patterns enhances adaptability to novel threats. For example, Pattern\_10 achieves an F1 score of 0.9977 in error-based detection, while Pattern\_4 reaches 0.9843 in network-based classification. This differentiation ensures that no single pattern dominates detection, leading to a more resilient classification system.

\begin{figure}[htp!]
    \centering
    \includegraphics[width=1\linewidth]{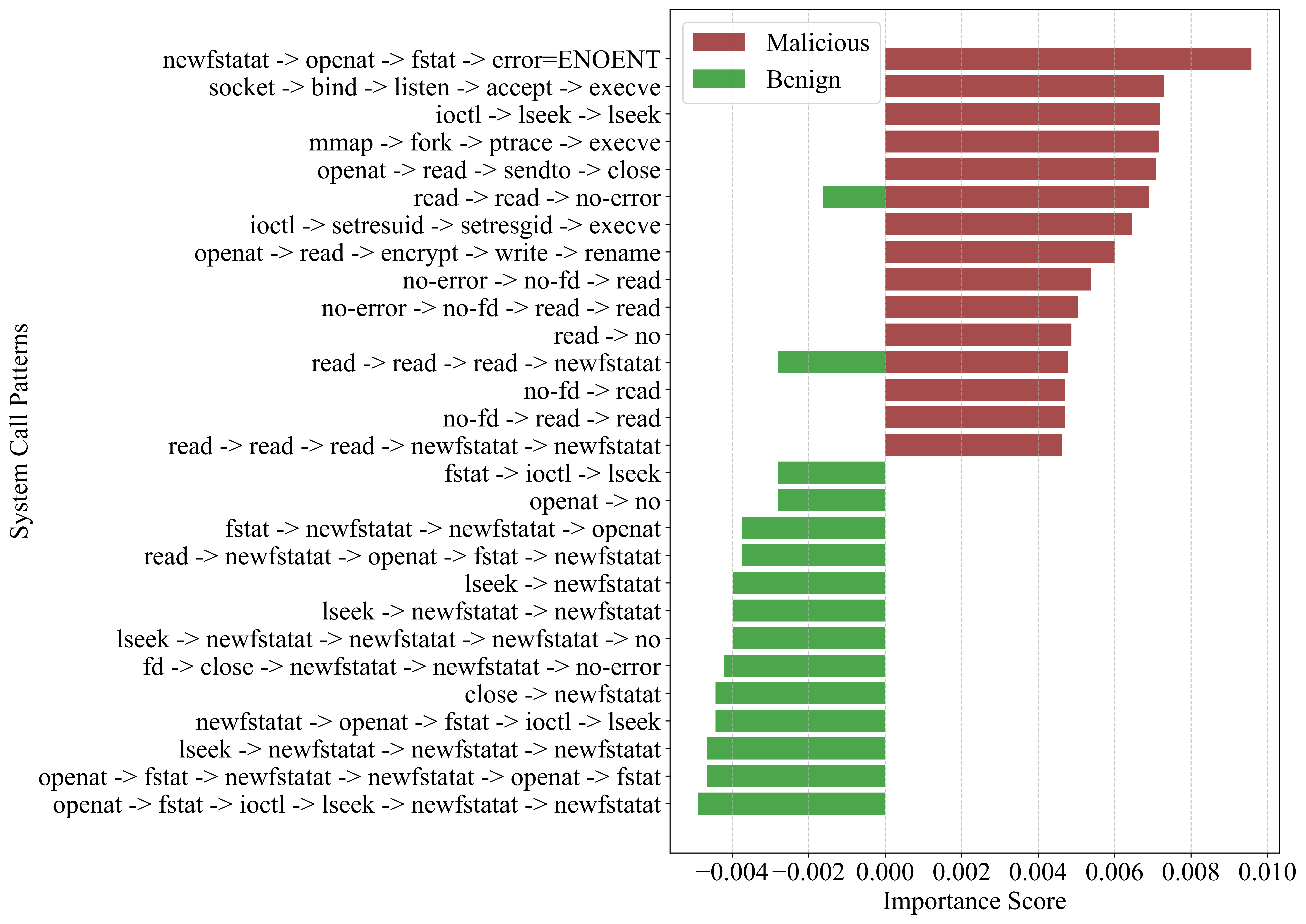}
    \caption{Top patterns contributing to classify malicious and benign package}
    \label{fig:Top patterns contributing}
\end{figure}

To further refine this classification, it is essential to identify which system call patterns are strongly indicative of malicious behavior and which are benign operations. Figure~\ref{fig:Top patterns contributing} illustrates this distinction by presenting high-impact system call sequences that serve as behavioral signatures for threat detection in DySec. Malicious activities are strongly linked to high-impact sequences such as \textit{newfstatat->openat->fstat->error=ENOENT} (probing for missing files) and \textit{socket->bind->listen->accept->execve} (network-based attacks), which reflect unauthorized access, error exploitation, or command-and-control infrastructure. Other notable malicious patterns include \textit{mmap->fork->ptrace->execve}, linked to dynamic code injection, and \textit{openat->read->encrypt->write->rename}, suggestive of ransomware-like operations. In contrast, benign patterns such as \textit{read->read->read->newfstatat->newfstatat} represent routine file operations, with moderate but consistent importance scores. These patterns reflect predictable workflows, such as metadata retrieval and file reads, typical of legitimate software operations. Also, benign packages often exhibit structured routines like \textit{fstat->ioctl->lseek}, which signify error-free installation workflows and standard input/output operations. 

While most system call patterns exhibit clear distinctions between benign and malicious behavior, some, such as \textit{read->read->read->newfstatat} and \textit{read->read->no-error}, appear in both contexts, introducing potential overlap. In benign scenarios, these patterns represent routine operations like dependency resolution and data retrieval, whereas in malicious cases, they may indicate reconnaissance or preparatory steps for data exfiltration, depending on their surrounding execution sequences. Despite this overlap, the structural independence of most patterns ensures that benign and malicious activities remain distinguishable. DySec effectively addresses this challenge by analyzing system call patterns within their broader execution context, enabling precise classification.

\begin{table*}[thb!]
\centering
\caption{Real-world detection analysis using 500 benign packages, and 500 malicious packages; bold indicates the best values.}
\label{tab: real-world}
\renewcommand{\arraystretch}{1}
\setlength{\tabcolsep}{3pt}
\centering
\begin{tabularx}{\textwidth}{|>{\centering\arraybackslash}c|>{\centering\arraybackslash}c|>{\centering\arraybackslash}X|>{\centering\arraybackslash}X|>{\centering\arraybackslash}X|>{\centering\arraybackslash}X|>{\centering\arraybackslash}X|>{\centering\arraybackslash}X|>{\centering\arraybackslash}X|>{\centering\arraybackslash}X|>{\centering\arraybackslash}c|}
\hline
\multirow{2}{*}{\textbf{Methods}} & \multirow{2}{*}{\textbf{Model}} & \multicolumn{6}{c|}{\textbf{Unseen Most Popular and Downloaded Packages (1,000)}} & \multirow{3}{*} \makecell{\textbf{Cmn. FP Package}} & \multirow{2}{*} \makecell{\textbf{Report on PyPI}} & \multirow{2}{*} \makecell{\textbf{PyPI Status}} \\ \cline{3-8}
 &  & \textbf{Precision} & \textbf{F1 Score} & \textbf{TP} ↑ & \textbf{TN} ↑ & \textbf{FP} ↓ & \textbf{FN} ↓ &  &  &  \\ \hline
\multirow{4}{*}{\makecell{Metadata \\Analysis~\cite{metadata_halder_2024}}} & RF & 0.9041 & 0.8536 & 452 & 393 & 48 & 107 & \multirow{4}{*}{44} & \multirow{4}{*}{No} & \multirow{4}{*}{-} \\  
 & DT & 0.8980 & 0.8376 & 449 & 377 & 51 & 123 & & &  \\  
 & SVM & 0.8440 & 0.7932 & 422 & 358 & 78 & 142 & & &  \\  
 & GB & 0.8819 & 0.8320 & 441 & 381 & 59 & 119 & & &  \\ \hline
\multirow{4}{*}{\makecell{Static \\Analysis~\cite{hybrid_samaana_2024}}} & RF & 0.9540 & 0.9111 & 477 & 430 & 23 & 70 & \multirow{4}{*}{17} & \multirow{4}{*}{No} & \multirow{4}{*}{-} \\  
 & DT & 0.9241 & 0.8791 & 462 & 411 & 38 & 89 & &  &  \\ 
 & SVM & 0.9259 & 0.8686 & 463 & 397 & 37 & 103 &  & & \\  
 & GB & 0.9380 & 0.8967 & 469 & 423 & 31 & 77 &  & &  \\ \hline
\multirow{4}{*}{\makecell{Dynamic \\Analysis [DySec]}} & RF & \textbf{0.9740} & \textbf{0.9681} & 487 & \textbf{481} & 13 & \textbf{19} & \multirow{4}{*}{11} & \multirow{4}{*}{\makecell{Report 6 \\Malicious \\Packages}} & \multirow{4}{*}{\makecell{Removed 4 \\Malicious \\Packages from \\PyPI}} \\  
 & DT & 0.9700 & 0.9528 & 485 & 467 & 15 & 33 &  &  &  \\ 
 & SVM & 0.9680 & 0.9434 & 484 & 458 & 16 & 42 & &  &  \\ 
 & GB & 0.9761 & 0.9606 & \textbf{488} & 472 & \textbf{12} & 28 & & &  \\ \hline
\end{tabularx}
\end{table*}

\vspace{-10pt}

\subsection{RQ3: Is DySec accurate enough to be practical?}

To evaluate the practical applicability of the proposed DySec framework, a real-world analysis was conducted on 1,000 PyPI packages (500 benign and 500 malicious). The benign samples were selected from the latest most popular and frequently downloaded PyPI packages, ensuring they had never been exposed to the detection model~\cite{pypistats_top_projects}. The malicious samples, sourced from recently disclosed threats identified on platforms like Fortinet, BleepingComputer, The Hacker News, ReversingLabs, and Kaspersky~\cite{fortinet2024, bleepingcomputer2024, kaspersky2024}, represent diverse attack patterns. To assess DySec’s effectiveness against emerging threats, we performed a second evaluation on packages updated between July and November 2024.

As shown in Table~\ref{tab: real-world}, the RF classifier within the DySec framework achieved the best performance, with a precision of 97.40\% and an F1-score of 96.81\%. It outperformed both metadata and static analysis methods. It also maintained low FP (1.3\%) and FN (1.9\%) rates, demonstrating its reliability in accurately detecting malicious packages while minimizing disruptions to benign workflows.

Notably, the RF classifier with CombinedTraces in DySec flagged 11 commonly used benign packages as potentially malicious during testing. Upon further investigation, six exhibited malicious behaviors, including data exfiltration, port scanning, socket proxy, and unauthorized remote access. All six were reported to PyPI maintainers with supporting dynamic analysis results. Four of these - `vermillion-0.5', `eth-abcde-0.2.3', `Pytonlib-0.0.0', and `infoind-3897' - were removed from PyPI maintainers following DySec’s findings. The remaining two packages - `PySocks-1.7.1' and `escposprinter-6.2' - were contested by PyPI maintainers. For `PySocks-1.7.1', maintainers clarified that the package is intentionally a socket proxy server and not inherently malicious. DySec’s detection of socket proxy behavior highlights its ability to flag functionalities that, while legitimate, could be repurposed for malicious activities such as anonymizing attack traffic. For `escposprinter-6.2', maintainers noted that the package bundles NetCat (nc.exe), a network utility tool, but asserted that the binary contains no recompiled malware. DySec’s alert on NetCat integration demonstrates its sensitivity to code patterns often abused in attacks, such as unauthorized network tool deployment. After validation, five additional false positive packages misclassified as malicious were confirmed benign, reaffirming DySec’s low false positive rate of 1.3\%. This outcome - successfully uncovering hidden threats while minimizing FN - illustrates DySec’s practical effectiveness in real-world scenarios.

\vspace{-5pt}

\section{Threats to Validity and Limitation}
\label{Threats to Validity}

\textbf{Internal validity:} The performance of the proposed DySec framework depends on the quality and representativeness of the collected dataset. Noise in install-time traces  (e.g., other IP/ port access during package installation) may impact its effectiveness by obscuring meaningful patterns. To mitigate these risks, all traces were collected in a secure, isolated Linux environment using eBPF technology. However, the reliance on a Linux-based setup introduces a limitation, as eBPF is exclusive to the Linux kernel. Additionally, configuring this isolated environment for dynamic analysis requires significant time and computational resources.

Another challenge arises when installing packages in a shared environment, as dependencies are automatically installed alongside primary packages. If these dependencies are reinstalled, the system may recognize them as already present, potentially distorting dependency resolution and analysis accuracy. To address this, each package was installed and processed in a dedicated virtual environment, minimizing dependency conflicts and ensuring consistency. Furthermore, the trace collection process generates large volumes of data, complicating feature extraction and ML model training. To address this, rigorous preprocessing steps, including data cleaning, data transformation, and feature extraction, were applied to ensure data integrity, thereby enhancing framework robustness.

\textbf{External validity:} This threat concerns the generalizability of findings to other ecosystems or package registries. While the framework was evaluated on a comprehensive dataset of PyPI packages (both benign and malicious), real-world scenarios may involve malware with unique characteristics not captured in the dataset. To partially address this, the framework was tested on 1,000 unseen, recently published popular packages. However, further validation across repositories such as NPM is necessary to confirm its applicability in diverse environments.

\textbf{Construct validity:} This threat stems from the reliance on dynamic features extracted from eBPF-based install-time traces. While these features effectively capture installation-phase behaviors, they may fail to detect delayed or runtime-specific malicious actions. Although the selected features cover a broad range of threats, sophisticated attackers employing advanced obfuscation techniques could still evade detection. Therefore, further exploration of runtime behavior monitoring is necessary to complement the install-time analysis and improve detection capabilities.

\vspace{-5pt}
\section{Conclusion and Future Works}
\label{Conclusion}

This study introduces DySec, an ML-based dynamic analysis framework designed to detect malicious packages within the PyPI ecosystem. It extracts features from install-time execution traces collected in a secure, isolated Linux environment using eBPF technology. To evaluate the approach, a comprehensive dataset comprising 36 dynamic, independent features was developed and tested using four ML classifiers. Among these, the RF classifier demonstrated the best performance, while the SVM classifier struggled to detect known malicious packages. DySec significantly enhances PyPI security compared to traditional metadata or static analysis methods by identifying unique malicious patterns and reducing exploitation risks. Furthermore, DySec demonstrated practical effectiveness, achieving overall low FN and FP rates when evaluated on unseen, widely used, and recently updated packages. These results underscore its potential for automated malware detection in real-world scenarios. Notably, DySec identified six packages previously classified as benign but exhibiting malicious behaviors, including data exfiltration, port scanning, and unauthorized remote access during installation. Four of these were subsequently removed by PyPI maintainers. These findings mark a critical advancement in securing open-source repositories, addressing software supply chain vulnerabilities, and fostering safer development ecosystems. In the future, we aim to bridge research and practice by developing a user-friendly platform for PyPI maintainers and developers. This platform will facilitate real-world testing, allowing users to upload package files for automated trace collection and ML-based malicious behavior prediction. Additionally, to enhance feature quality, advanced techniques such as multi-modal analysis and domain-specific behavioral modeling will be explored to detect malicious patterns more effectively.

\bibliographystyle{IEEEtran}
\bibliography{refrences}

\begin{figure*}
    \centering
    \includegraphics[width=0.90\linewidth]{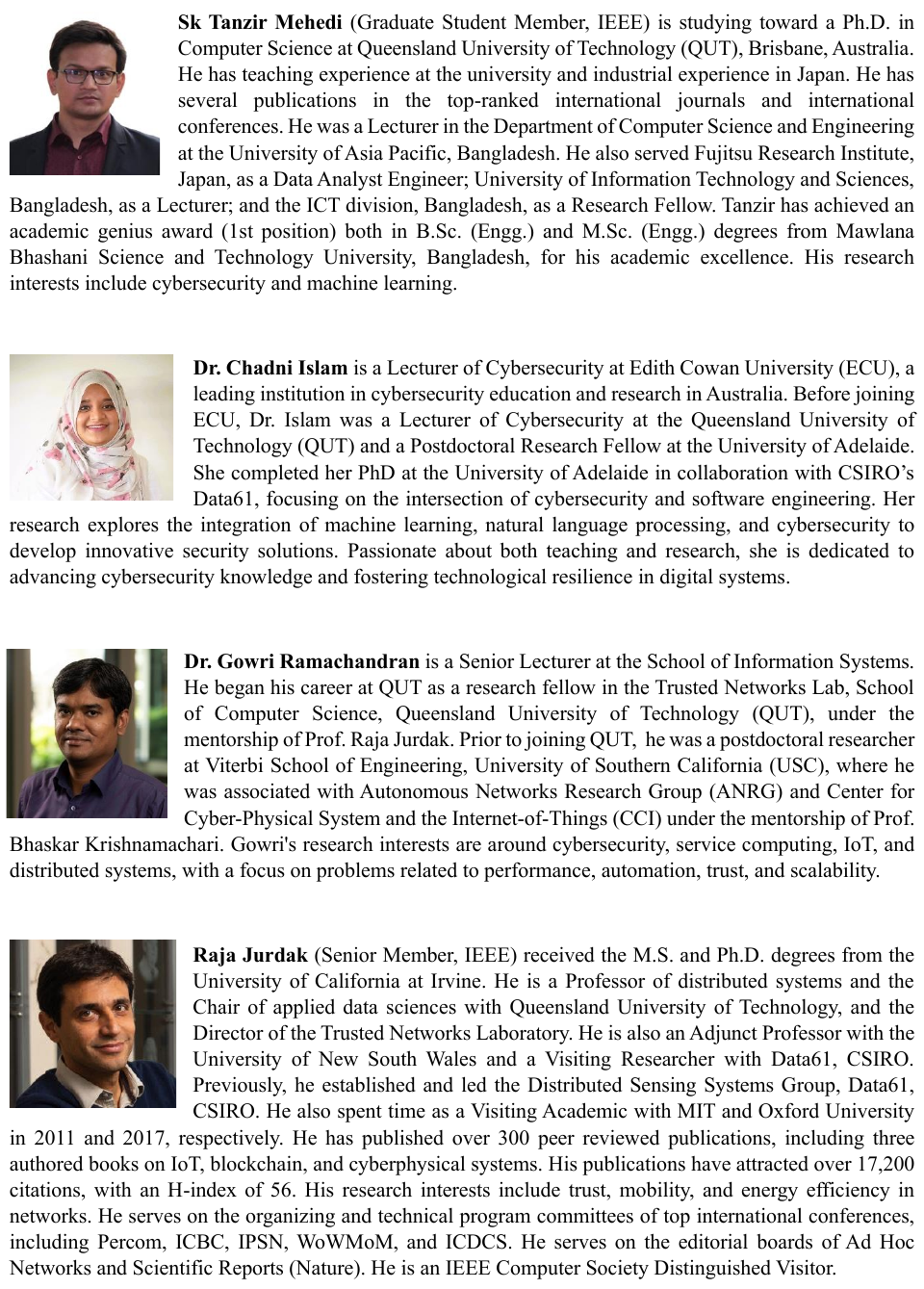}
\end{figure*}

\begin{figure*}
    \centering
    \includegraphics[width=1\linewidth]{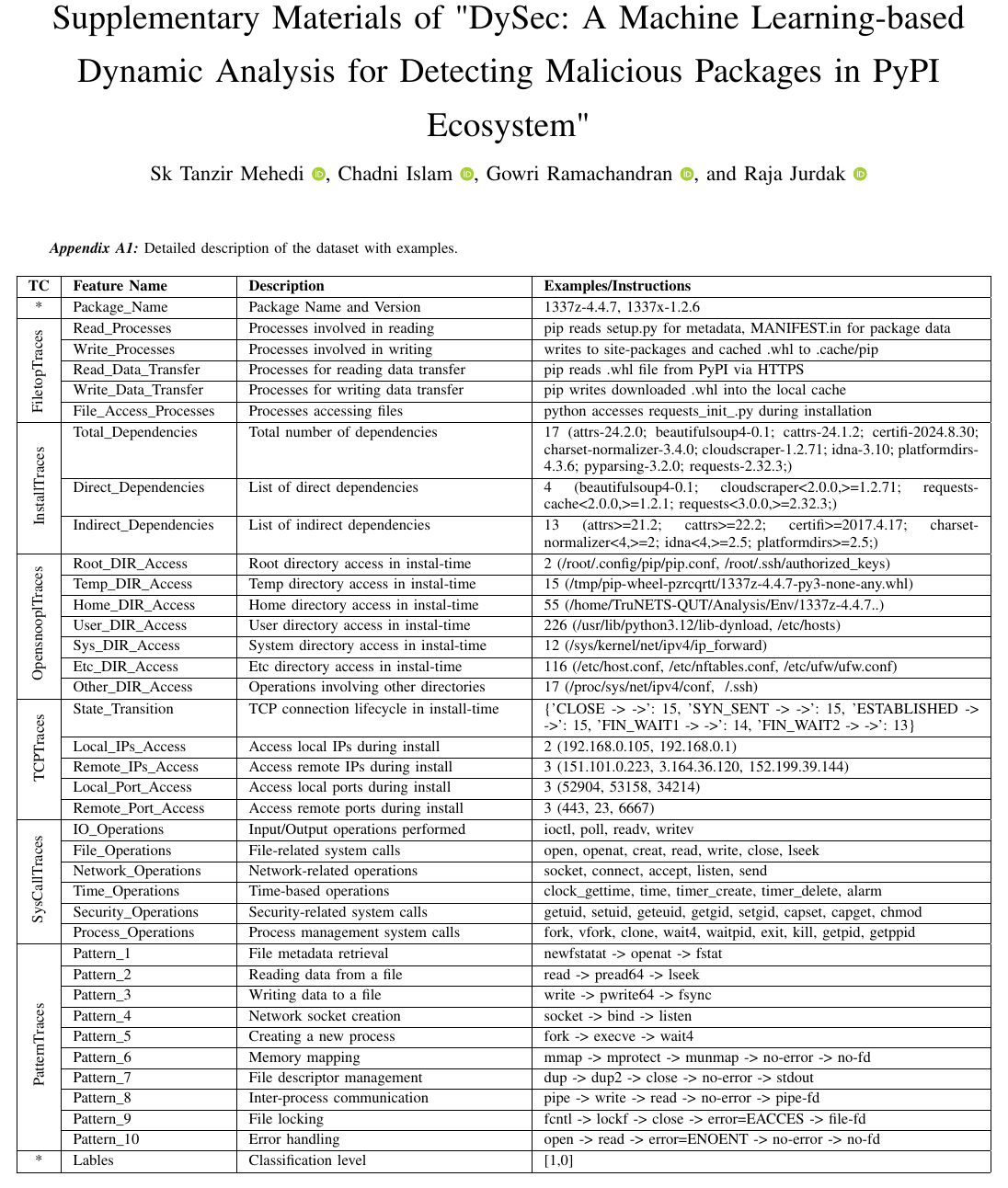}
\end{figure*}

\end{document}